\documentclass[12pt]{article}

\usepackage{graphicx}
\usepackage{amsfonts}
\usepackage{amsthm}
\usepackage{amsmath}
\usepackage[top=3cm, bottom=3.25cm, left=2.75cm, right=2.75cm]{geometry}

\newcommand{\beq}{\begin{equation}}
\newcommand{\eeq}[1]{\label{#1}\end{equation}}
\newcommand{\ber}{\begin{eqnarray}}
\newcommand{\eer}[1]{\label{#1}\end{eqnarray}}
\newcommand{\re}[1]{(\ref{#1})}
\newcommand{\ie}{{i.e.},~}
\numberwithin{equation}{section}

\newcommand{\bbD}[1]{\mathbb{D}_{#1}}
\newcommand{\bbDB}[1]{\bar{\mathbb{D}}_{#1}}
\newcommand{\bbnab}{\mbox{\hbox{$\nabla$\kern-0.65em\lower.39ex\hbox{${}^{\nabla}$}}}}

\newcommand{\bbX}[1]{\mathbb{X}^{#1}}
\newcommand{\bbXB}[1]{\bar{\mathbb{X}}^{#1}}

\newcommand{\hcd}[1]{\nabla_{#1}}

\newcommand{\fb}{{\bf f}}
\newcommand{\fbb}{{\bf \bar f}}

\def\one{{1\!\! 1}}
\def\+{{+\!\!\!+}}
\def\pp{\mbox{\tiny${}_{\stackrel\+ =}$}}

\newcommand{\kah}{K\"ahler~}

\newcommand{\eg}{{e.g.},~}
\newcommand{\half}{\frac 1 2}

\begin{document}
\renewcommand{\theequation}{\thesection.\arabic{equation}}
\setcounter{page}{0}
\thispagestyle{empty}
\begin{flushright} \small
UUITP-28/09 \\
YITP-SB-09-42\\
\end{flushright}

\smallskip
\begin{center}
 \LARGE
{\bf Sigma models with off-shell ${ N}=(4,4) $ supersymmetry and noncommuting complex structures}
\\[12mm]
 \normalsize
{\bf M.~G\"oteman${}^a$, U.~Lindstr\"om${}^a$, M.~Ro\v cek${}^b$, and Itai Ryb${}^{ab}$} \\[8mm]
 {\small\it
a: Department Physics and Astronomy,\\
Division for Theoretical Physics,\\
Uppsala University, \\ Box 803, SE-751 08 Uppsala, Sweden
\bigskip

b: C.N. Yang Institute for Theoretical Physics,\\
Stony Brook University,\\
Stony Brook, NY 11794-3840, USA}
\end{center}
\vspace{10mm} \centerline{\bfseries Abstract} \bigskip

\noindent We describe the conditions for extra supersymmetry in $N=(2,2)$ 
supersymmetric nonlinear sigma models
written in terms of semichiral superfields. We find that some of these models 
have additional off-shell supersymmetry. The $(4,4)$ supersymmetry 
introduces geometrical structures on the target-space which are conveniently 
described in terms of Yano $f$-structures and Magri-Morosi concomitants. 
On-shell, we relate the new structures to the known bi-hypercomplex structures.
\eject
\tableofcontents


\section{Introduction}
The target-space geometry of two-dimensional supersymmetric nonlinear 
sigma models has been extensively discussed in the literature.
In \cite{Gates:1984nk}, and partly in \cite{Howe:1985pm}, the general case 
including a $B$-field was described in $(1,1)$ superspace. For $(2,2)$ 
supersymmetry the target-space geometry was shown to be 
bihermitean, \ie the metric is hermitean with respect to two complex structures 
$J_{(\pm)}$. Off-shell, a manifest $(2,2)$ formulation was only found 
when the complex structures commute\footnote{Some other models with off-shell (2,2) supersymmetry
were found in \cite{Buscher:1987uw}--\cite{Bogaerts:1999jc}.}. 
Similar results hold for $(4,4)$ supersymmetry: A manifest $(2,2)$ 
formulation was only found when (some of) the complex structures commute.

More recently the bihermitean geometry of \cite{Gates:1984nk} has been described 
as generalized K\"ahler geometry \cite{Gualtieri:2003dx}, a subclass of generalized 
complex geometry \cite{{hitchinCY}}. The intimate relation of this description to 
sigma models is elucidated in, \eg\cite{Lindstrom:2004eh}--\cite{Lindstrom:2007qf}. 
In particular, as shown in \cite{Lindstrom:2005zr}, a complete $(2,2)$ superspace 
description of generalized 
K\"ahler geometry, including the case when the complex structures do not commute, 
requires semichiral fields \cite{Buscher:1987uw} in addition to the chiral and twisted 
chiral fields; this had been conjectured but not proven by
Sevrin and Troost \cite{Sevrin:1996jr}. The superspace lagrangian $K$ is further 
shown to be a potential for the metric and $B$-field \cite{Lindstrom:2007xv};

The bi-hypercomplex geometry of \cite{Gates:1984nk} has likewise been described 
as generalized hyperk\"ahler geometry in \cite{Bredthauer:2006sz} and \cite{Bobby}.
\bigskip

In the present paper, we discuss models written in terms of semichiral
fields only. We ask under which conditions such a model can carry
$(4,4)$ supersymmetry. A limited class of such models was recently
discussed in \cite{Goteman:2009xb}. There the extra
transformations were taken to be linear in the derivatives of the
fields, and the target-space was restricted to be four-dimensional. It
was found that no interesting solution for $N=(4,4)$ supersymmetry exists, but instead one can find an interesting solution for $N=(4,4)$ twisted supersymmetry. This implied that the target-space must have pseudo-hypercomplex geometry.

Some models including semichiral but no chiral or twisted chiral
fields had been treated previously in \cite{Lindstrom:1994mw}; they
include additional auxiliary $(4,4)$ fields, and only become purely
semichiral models on-shell.

Models with commuting complex structures, described by $n$ chiral and
$m$ twisted chiral fields, have off-shell $(4,4)$ supersymmetry when
$n=m$ and the Lagrangian $K$ satisfies certain differential
constraints \cite{Gates:1984nk}. Purely semichiral models have to have
an equal number of left and right semichiral fields
\cite{Buscher:1987uw}. Here we find that for some such models whose
Lagrangian again satisfies certain differential constraints, there is
an off-shell algebra. This algebra has an interpretation in terms of
an integrable Yano $f$-structure on $TM\oplus TM$, the sum of two
copies of the tangent bundle of the target-space. We already know from
\cite{Gates:1984nk} that a sigma model with $(4,4)$ supersymmetry has
two quaternion-worth of complex structures, $J_{(\pm)}^ A$, living on
$TM$ and we find that all of these structures fit together nicely. In
particular we resolve the interplay between the various integrability
conditions involving Nijenhuis tensors and Magri-Morosi concomitants.

The generalized K\"ahler potential for those semichiral models 
that are invariant under the off-shell algebra satisfy a constraint.
This is analogous to the $(4,4)$ conditions in \cite{Gates:1984nk} 
which are realized for commuting complex structures by the $N=4$
twisted chiral multiplet. For a subclass of our models, 
we can give a geometric interpretation of the condition as a kind of 
hermiticity condition: a certain tensor is preserved by the $f$-structures.

We follow the method used in previous discussions of additional
nonmanifest supersymmetries, \eg in \cite{Gates:1984nk} and
\cite{Hull:1985pq}. To study the additional symmetries, we make the
most general ansatz compatible with the properties of the superfields,
and then read off the constraints that follow from closure of the
supersymmetry algebra and invariance of the action. The constraints
from the algebra are discussed in section \ref{nonman}, the invariance
of the action is presented in section \ref{inv}.  Often in these
investigations field-equations arise and the algebra only closes
on-shell. In section \ref{offshell} we analyze off-shell closure while
postponing the on-shell discussion to section \ref{OnShell}.

\section{Preliminaries}
\label{prelim}

This section contains background material needed for the discussions in later sections.

The $(2,2)$ supersymmetry algebra for the covariant derivatives is given by
\beq
\{\bbD{\pm},\bbDB{\pm}\}=i\partial_{\pp}~,
\eeq{alg}
and the left and right semichiral fields $\bbX{a,a'}$, and left and right 
anti-semichiral fields $\bbXB{\bar a,\bar a'}$ \cite{Buscher:1987uw}
satisfy 
\beq
\bbDB{+}\bbX{a}=0~, \quad \bbDB{-}\bbX{a'}=0, \quad
\bbD{+}\bbXB{\bar a}=0~, \quad \bbD{-}\bbXB{\bar a'}=0~.
\eeq{000}
A useful collective notation, often used in previous papers, is 
$\bbX{L}=(\bbX{a},\bbXB{\bar a})$ and \\$\bbX{R}=(\bbX{a'},\bbXB{\bar a'})$. 
When we need a notation for all of the fields we write $\bbX{i}$ with $i=(L,R)$.

We shall consider the generalized \kah potential $K$
and the sigma model it defines through the action
\beq
S=\int d^2\xi \bbD{}^2 \bbDB{}^2 ~\!K(\bbX{i})~.
\eeq{act1}
 The target-space manifold ${\cal M}^{4d}$ coordinatized by the $d$ 
 left and $d$ right semichiral fields (and their conjugates) carries bihermitean 
 geometry. This means that there are two complex structures $J_{(\pm)}$, a
 metric $g$ hermitean with respect to both of these and a closed three 
 form $H$ such that \cite{Gates:1984nk}
 \ber\nonumber
 &&J_{(\pm)}^2=-\one\\[1mm]\nonumber
 && \nabla^{(\pm)}J_{(\pm)}=0, \quad \Gamma^{(\pm)}=
 \Gamma^0 \pm \half g^{-1}H\\[1mm]\nonumber
&&J_{(\pm)}^tgJ_{(\pm)}=g~\\[1mm]
&& d^c_+ \omega_++d^c_- \omega_-=0, \quad H=d^c_+ \omega_+=-d^c_- \omega_-~,
 \eer{cs}
 where $\Gamma^0$ is the Levi-Civita connection for the metric $g$ and
 $
 d_{(\pm)}^c:= J_{(\pm)}(d)~.
 $
 The expression for $d^c$ becomes most simple in complex 
 coordinates: $d^c= i(\bar\partial-\partial)$. 
 
In later sections we shall also need the explicit form of the complex structures:
They are defined in terms of the matrices \cite{Lindstrom:2005zr}
\beq
K_{LR}:=\left(\begin{array}{cc}K_{aa'}&K_{a\bar a'}\cr 
K_{\bar a a'}&K_{\bar a\bar a'}\end{array}\right)~,
\eeq{KLR}
and with
$C:=[j,K],$
they read
\ber
&&J_{(+)}=\left(\begin{array}{cc}
j&0\cr
K^{-1}_{RL}C_{LL}&K^{-1}_{RL}jK_{LR}
\end{array}\right)\cr
&&~~\cr
&&~~\cr
&&J_{(-)}=\left(\begin{array}{cc}
K^{-1}_{LR}jK_{RL}&K^{-1}_{RL}C_{RR}\cr
0&j
\end{array}\right)
\eer{jpm}
with $j$ denoting a canonical $2d\times 2d$ complex structure
\beq
j:=\left(\begin{array}{cc}
i&0\cr
0&-i\end{array}\right)~.
\eeq{jjj}
 
The description \re{cs} applies to bihermitean geometry in general,
which may be described using chiral, twisted chiral and semichiral
fields \cite{Lindstrom:2005zr}. A special feature of the case we are
interested in here is that, although locally we may always write
$H=dB$, for the model with only semichiral fields $B$ is globally
defined (away from type change loci \cite{Gualtieri:2003dx}). For more
aspects of the global structure of bihermitean geometry, see
\cite{Hull:2008vw}.

The data $(g,B,J_{(\pm)})$ in (\ref{cs}) may be packaged as structures
on $TM\oplus T^*\! M$ in the form of generalized K\"ahler geometry
\cite{Gualtieri:2003dx}.
 
\section{Nonmanifest supersymmetries}
\label{nonman}

\subsection{Ansatz for non-manifest supersymmetries}
Requiring that the derivatives are covariant with respect to the 
additional supersymmetries, \eg $\bbDB{+}(\delta \bbX{a})=
\delta(\bbDB{+}\bbX{a})=0$, leads to the following general 
ansatz for $N=(4,4)$ supersymmetry:
\ber\nonumber
\delta \bbX{a}&=&
\bar\epsilon^+\bbDB{+}f^a(\bbX{L,R}, \bbXB{L,R})
+g^a_b(\bbX{c})\bar\epsilon^-\bbDB{-}\bbX{b}+h^a_b(\bbX{c})
\epsilon^-\bbD{-}\bbX{b}~,\\[1mm] \nonumber
\delta \bbXB{\bar a}&=&\epsilon^+\bbD{+}\bar f^{\bar a}
(\bbX{L,R}, \bbXB{L,R}) + \bar g^{\bar a}_{\bar b} (\bbXB{\bar c})\epsilon^-\bbD{-}\bbXB{\bar b}
+\bar h^{\bar a}_{\bar b}(\bbXB{\bar c}) \bar\epsilon^-\bbDB{-}\bbXB{\bar b}~,\\[1mm]\nonumber
\delta \bbX{a'}&=&\bar\epsilon^-\bbDB{-}\tilde f^{a'}
(\bbX{L,R}, \bbXB{L,R})+\tilde g^{a'}_{b'}(\bbX{c'})
\bar \epsilon^+\bbDB{+}\bbX{b'}+\tilde h^{a'}_{b'}(\bbX{c'})
\epsilon^+\bbD{+}\bbX{b'}~,\\[1mm]
\delta \bbXB{\bar a'}&=& \epsilon^-\bbD{-}\bar{\tilde f}^{\bar a'} (\bbX{L,R}, \bbXB{L,R})
+ \tilde {\bar g}^{\bar a'}_{\bar b'}(\bbXB{\bar c'}) \epsilon^+\bbD{+}\bbXB{\bar b'}
+ \bar{\tilde h}^{\bar a'}_{\bar b'}(\bbXB{\bar c'}) \bar\epsilon^+\bbDB{+}\bbXB{\bar b'}~,
\eer{2}
where $\epsilon^\pm$ are the transformation parameters. 
This ansatz is covariant under left and right holomorphic transformations,  
i.e., coordinate transformations of the form\footnote{Strictly speaking, 
these are not the most general left and right holomorphic transformations, 
as they also preserve the choice of polarization, i.e., the separation into left and right coordinates.}
\ber\nonumber
&&\bbX{a}\to\bbX{'a}(\bbX{b})~~,~~~ 
\bbX{\bar a}\to\bbX{'\bar a}(\bbX{\bar b}) ~,\\[1mm]
&&\bbX{a'}\to\bbX{'a'}(\bbX{b'})~~,~~~ 
\bbX{\bar a'}\to\bbX{'\bar a'}(\bbX{\bar b'})~. 
\eer{holo}
A useful way of rewriting these nonmanifest transformations 
introduces the matrices $U^{(\pm)}$ and $V^{(\pm)}$ defined as
\ber
&\bar\delta^\pm\bbX{}:=\bar\delta^\pm\left( \begin{array}{l}\bbX{a}\cr
\bbXB{\bar a}\cr\bbX{a'}\cr
\bbXB{\bar a'}\end{array}\right)=\bar\delta^\pm\left( \begin{array}{l}\bbX{L}\cr\bbX{R}\cr\end{array}\right)=
U^{(\pm)}\bar\epsilon^\pm\bbDB{\pm}\bbX{}~,
\quad
&\delta^\pm\bbX{}=
V^{(\pm)}\epsilon^\pm\bbD{\pm}\bbX{}\
\eer{MNdef}
where\footnote{The fundamental tensorial objects are defined in (\ref{2}). 
Additional covariant indices denote partial derivatives, \eg $f^a_i:=\partial_if^a$,  etc.}
\beq
U^{(+)}=\left( \begin{array}{cccc}
\ast &f^{a}_{\bar b}&f^{a}_{ b'}&f^{a}_{\bar b'}\cr
\ast &0&0&0\cr
\ast &0&\tilde g^{a'}_{b'}&0\cr
\ast &0&0&\bar{\tilde h}^{\bar a'}_{\bar b'}\end{array}\right),
\quad
U^{(-)}=\left( \begin{array}{cccc}
g^a_b & 0 &\ast &0\cr
0 &\bar h^{\bar a}_{\bar b} &\ast &0\cr
\tilde f^{a'}_{ b} &\tilde f^{a'}_{\bar b} &\ast &\tilde f^{a'}_{\bar b'}\cr
0&0&\ast &0\end{array}\right)
\eeq{MNdef2}
and
\beq
V^{(\pm)}=\left(\begin{array}{cc} \sigma_1 & 0\cr
0 &\sigma_1\end{array}\right)\bar U^{(\pm)}\left(\begin{array}{cc} \sigma_1 & 0\cr
0 &\sigma_1\end{array}\right).
\eeq{Ndef}
Here
\beq \sigma_1=\left(\begin{array}{cc}0&1\cr
1&0\end{array}\right)~.
\eeq{Pauli1} 
Note that one column in each of the transformation matrices
$U^{(\pm)}$ and $V^{(\pm)}$ is arbitrary. For the remainder of the
paper, we set the arbitrary entries to zero. Doing so provides us with
full integrability of the transformation matrices and an
interpretation of the off-shell algebra in terms of Yano
$f$-structures. The consequences of keeping the arbitrariness is
discussed briefly in section \ref{disk}.

For later use, we introduce the projection operators $P^\pm,\hat P^\pm$:
\ber
P_+=\left(\begin{array}{cccc}0 & 0 & 0 & 0 \\0 & 1 & 0 & 0 
\\0 & 0 & 0 & 0 \\0 & 0 & 0 & 0\end{array}\right)~&,&~~
\hat P_+=\left(\begin{array}{cccc}1 & 0 & 0 & 0 \\0 & 0 & 0 & 0 
\\0 & 0 & 0 & 0 \\0 & 0 & 0 & 0\end{array}\right)~,~~\nonumber \\[5mm]
P_-=\left(\begin{array}{cccc}0 & 0 & 0 & 0 \\0 & 0 & 0 & 0 
\\0 & 0 & 0 & 0 \\0 & 0 & 0 & 1\end{array}\right)~&,&~~
\hat P_-=\left(\begin{array}{cccc}0 & 0 & 0 & 0 \\0 & 0 & 0 & 0 
\\0 & 0 & 1 & 0 \\0 & 0 & 0 & 0\end{array}\right)~.
\eer{projectors}
\subsection{Magri-Morosi concomitant}
To interpret the expressions we find below, we use the Magri-Morosi concomitant 
\cite{YanoAko,MagriMorosi} 
defined for two endomorphisms $I$ and $J$ of the tangent bundle $TM$ 
of a manifold $M$ as 
\begin{equation}
\mathcal M(I,J)^i_{jk} :=-
\mathcal M(J,I)^i_{kj}= I^l{}_j J^i{}_{k,l}- J^l{}_k I^i{}_{j,l} - I^i{}_l J^l{}_{k,j} + J^i{}_l I^l{}_{j,k}~. 
\end{equation}
This concomitant has previously been used when discussing
supersymmetry algebra, \eg in discussing $(1,0)$ and $(1,1)$
formulations of certain $(p,q)$ sigma models in \cite{Howe:1988cj} and
discussing generalized complex geometry for $(2,2)$ models in
\cite{Bredthauer:2006hf}.

The Magri-Morosi concomitant relates to the simultaneous integrability
of two structures and is a tensor only when $[I,J]=0$. More precisely,
two commuting complex structures are simultaneously integrable if and
only if  their Magri-Morosi concomitant vanishes. The part
antisymmetric in $j,k$ is the Nijenhuis concomitant ${\cal N}(I,J)$;
when $I=J$ this becomes the Nijenhuis tensor ${\cal N}(I)$. If ${\cal
N}(I)=0$, then $I$ is integrable.

Assuming that we have one $I$-connection $\nabla^{(I)}$ and one
$J$-connection $\nabla^{(J)}$ differing only in the sign of the
torsion $\Gamma^{(I/J)}=\Gamma^{(0)}\pm T$, we can rewrite $\mathcal
M$ as
\ber\nonumber
\mathcal M(I,J)^i_{jk} &=& I^l_{~j} 
\hcd{l}^{(J)}J^i_{~k}-J^l_{~k} \hcd{l}^{(I)}I^i_{~j} - I^i_{~l} 
\hcd{j}^{(J)}J^l_{~k} + J^i_{~l} \hcd{k}^{(I)}I^l_{~j}
-[I,J]^i_m\Gamma_{jk}^{(J)~m}\\[1mm]
&:=&\widehat {\mathcal M}(I,J)^i_{jk} -[I,J]^i_m\Gamma_{jk}^{(J)~m}~.
\eer{plusminuscon}
We shall need this version in section \ref{onshell} below.

Finally, we note that in the special case when $I^i_{~j}$ and $J^i_{~j}$ 
are curl-free in the lower indices, the concomitant simplifies to
\beq
{\mathcal{M}}(I,J)^i_{jk}=\left(JI\right)^i_{j,k}-\left(IJ\right)^i_{k,j}~.
\eeq{specon}

\subsection{Constraints from the supersymmetry algebra}
\label{algcstrts}

Imposing the left-with-right commutator algebra for the ansatz \re{MNdef} 
relates the Magri-Morosi concomitant of transformation matrices to the 
commutator of the same matrices as follows
\ber
[ \bar{\delta}^\pm,\bar{\delta}^\mp ]\mathbb{X}^i = 0 & 
\Longleftrightarrow &
{\mathcal{M}}(U^{(\pm)},U^{(\mp)})^i_{jk} 
\bar{\mathbb{D}}_{\pm}\mathbb{X}^j 
\bar{\mathbb{D}}_{\mp}\mathbb{X}^k  =
[U^{(\pm)},U^{(\mp)}]^i{}_m\bbDB{\pm} \bbDB{\mp}\bbX{m}~,\nonumber \\
{}[ \bar{\delta}^\pm,{\delta}^\mp ]\mathbb{X}^i = 0 & \Longleftrightarrow &
{\mathcal{M}}(U^{(\pm)},V^{(\mp)})^i_{jk} 
\bar{\mathbb{D}}_{\pm}\mathbb{X}^j{\mathbb{D}}_{\mp}
\mathbb{X}^k  =[U^{(\pm)},V^{(\mp)}]^i{}_m\bbDB{\pm} \bbD{\mp}\bbX{m}~.
\eer{noncov_closure1}
These relations can be rewritten covariantly using 
$\widehat{\mathcal M}$ defined in (\ref{plusminuscon}) as
\ber\nonumber
\widehat{\mathcal{M}}(U^{(\pm)},U^{(\mp)})^i_{jk} 
\bar{\mathbb{D}}_{\pm}\mathbb{X}^j \bar{\mathbb{D}}_{\mp}
\mathbb{X}^k & =& [U^{(\pm)},U^{(\mp)}]^i{}_m 
\left(\bar{\mathbb{D}}_{\pm}\bar{\mathbb{D}}_{\mp}\mathbb{X}^m + 
\Gamma_{jk}^{(\mp)~m}\bar{\mathbb{D}}_{\pm}
\mathbb{X}^j \bar{\mathbb{D}}_{\mp}\mathbb{X}^k \right)\\
& =&[U^{(\pm)},U^{(\mp)}]^i{}_m\bar{\bbnab}^{(\mp)}_\pm 
\bbDB{\mp}\bbX{m}~,\nonumber\\
\widehat{\mathcal{M}}(U^{(\pm)},V^{(\mp)})^i_{jk} 
\bar{\mathbb{D}}_{\pm}\mathbb{X}^j \mathbb{D}_{\mp}
\mathbb{X}^k & =& [U^{(\pm)},V^{(\mp)}]^i{}_m 
\left(\bar{\mathbb{D}}_{\pm}\mathbb{D}_{\mp}\mathbb{X}^m + 
\Gamma_{jk}^{(\mp)~m}\bar{\mathbb{D}}_{\pm}
\mathbb{X}^j \mathbb{D}_{\mp}\mathbb{X}^k\right)\nonumber \\
& =&[U^{(\pm)},V^{(\mp)}]^i{}_m
\bar{\bbnab}^{(\mp)}_\pm\bbD{\mp}\bbX{m}~.
\eer{list1}
In the last equalities we have identified the pullback of the 
covariant derivative, for use in the on-shell section. Note 
that constraints on the semichiral fields imply that some of the equations vanish trivially.

The constraints from the left-with-left and right-with-right 
part of the algebra involve the Nijenhuis tensor:
\beq
[\bar{\delta}^\pm,\bar{\delta}^\pm]\mathbb{X}^i= 0 \ 
\Longleftrightarrow\  \mathcal{N}(U^{(\pm)})^i_{jk} 
\bar{\mathbb{D}}_{\pm}\mathbb{X}^j\bar{\mathbb{D}}_{\pm}\mathbb{X}^k = 0~.
 \eeq{leftright}
Finally, using the algebra (\ref{alg}), the commutator 
$[\delta^\pm,\bar{\delta}^\pm]\mathbb{X}^i= i 
\bar{\epsilon}^\pm \epsilon^\pm \partial_{\pp}\mathbb{X}^i$
yields 
\ber
\mathcal{M}(U^{(\pm)},V^{(\pm)})^i_{jk} 
\bar{\mathbb{D}}_{\pm}\mathbb{X}^j \mathbb{D}_{\pm}\mathbb{X}^k \nonumber 
&=&\phantom{+}\left[\left(UV\right)^{(\pm)i}_{~j}+\delta^i_j
\right]\bbDB{\pm}\bbD{\pm}\bbX{j} \nonumber\\ 
&& +\left[\left(VU\right)^{(\pm)i}_{~j}+\delta^i_j\right]\bbD{\pm}\bbDB{\pm}\bbX{j}~.
\eer{bub1_noncov}

\section{Off-shell interpretation of the algebra constraints}
\label{offshell}

In this section we analyze the constraints found in section \ref{algcstrts}, 
separating the conditions into algebraically independent parts.
\subsection{The conditions for off-shell invariance}
Off-shell, $\mathbb{D}\mathbb{X}\mathbb{D}\mathbb{X}$ and 
$\mathbb{D}\mathbb{D}\mathbb{X}$ are independent structures 
and hence both sides in equation (\ref{noncov_closure1}) and 
\re{bub1_noncov} must vanish independently. This gives the conditions
\ber
\mathcal{M}(U^{(+)},U^{(-)})^i_{jk}\ =&0~, \quad &~~ j\ne a,~k\ne a'\cr
\mathcal{M}(U^{(+)},V^{(-)})^i_{jk}\ =&0~,\quad &~~ j\ne a,~k\ne \bar a'\cr
\mathcal{M}(U^{(+)},V^{(+)})^i_{jk}\ =&0~, \quad&~~ j \ne a,~ k \ne \bar a~,
\eer{MMzero_noncov}
and
\ber
[U^{(+)},U^{(-)}]^i{}_j\ =&0~,\quad &~~ j\ne a,a'\cr
[U^{(+)},V^{(-)}]^i{}_j\ = &0~,\quad &~~ j\ne a,\bar a'~,
\eer{sum}

and finally
\beq
\left(UV\right)^{(+)i}_{~j}\ =-\delta^i_j~, \quad j\ne \bar a~,\quad
\left(VU\right)^{(+)i}_{~j}\ =-\delta^i_j~, \quad j\ne a~,
\eeq{sumer1}
together with their complex conjugate equations.
Setting the arbitrary entries in the transformation matrices to 
zero sets the undetermined columns in (\ref{sumer1}) to zero,
\beq
\left(UV\right)^{(+)i}_{~\bar a}\ = 
\left(VU\right)^{(+)i}_{~a}\ =
\left(UV\right)^{(-)i}_{~\bar a'} =
\left(VU\right)^{(-)i}_{~a'}\ = 0~.
\eeq{sumer1c}
 
The constraint \re{leftright}
implies that $U^{(\pm)}$ and $V^{(\pm)}$ are 
integrable on some subspace. 
When we impose \re{sumer1c}, the 
integrability extends to the full space:
\beq
 \mathcal{N}(U^{(\pm)})^i_{jk} = 0~.
\eeq{Nije}
The conditions in \re{MMzero_noncov} may be 
written as in \re{specon} plus curl terms;
\beq
\mathcal{M}(U^{(\pm)},V^{(\pm)})^i_{kj}=
\left(VU\right)^{(\pm)i}_{~j,k}-
\left(UV\right)^{(\pm)i}_{k,j}+U^{(\pm)l}_{~j}
V^{(\pm)i}_{[k,l]}-V^{(\pm)l}_kU^{(\pm)j}_{[j,l]}=0~.
\eeq{sumerien}
The first two terms vanish due to \re{sumer1}. The form of the ansatz
\re{MNdef2} reveals that most of the third and fourth terms 
also vanish identically. The remaining ones may be shown to 
be zero due to \re{sumer1} and the integrability \re{Nije}. 
As an example of the last statement consider
\beq
U^{(+)l}_{~j}V^{(+)a'}_{[k,l]}-V^{(+)l}_kU^{(+)a'}_{[j,l]}
\eeq{explain}
which is nonvanishing for $j,k=b',d'$ when it becomes
\beq
\tilde h^{a'}_{[b',c']}\tilde g ^{c'}_{d'}-\tilde g^{a'}_{[d',c']}\tilde h ^{c'}_{b'}~.
\eeq{explainex}
 A short calculation then shows that this combination 
 is zero due to \re{sumer1} 
 \beq
 \tilde h^{a'}_{c'} \tilde g^{c'}_{d'}= \tilde g^{a'}_{c'} 
 \tilde h^{c'}_{d'}=-\delta^{a'}_{d'}~,
 \eeq{gh1}
 and \re{Nije}
 \beq
 \tilde g^{c'}_{[d'} \tilde g^{a'}_{b'],c'}-\tilde g^{a'}_{c'} 
 \tilde g^{c'}_{[b',d']}=0~. 
 \eeq{gg}

In summary, off-shell we find the following algebraic constraints 
in all sectors not projected out by the semi-chiral constraints:
\begin{itemize}
\item
The transformation matrices $U^{(\pm)}, V^{(\pm)}$ all commute.
\item
The products $U^{(\pm)}V^{(\pm)}$ and 
$V^{(\pm)}U^{(\pm)}$ equal minus one.
\item
The transformation matrices are all separately integrable.
\item
The Magri-Morosi concomitant vanishes for all two pairs of the 
transformation matrices. We showed that some of these, 
namely the last one in (\ref{MMzero_noncov}) relating $U^{(\pm)}$ 
with $V^{(\pm)}$, follow from the above three constraints.
\end{itemize}
The zeros in the arbitrary columns of the transformation matrices 
gives full integrability as in (\ref{Nije}) and the relations (\ref{sumer1c}). 
This makes the products $U^{(\pm)}V^{(\pm)}$ and $V^{(\pm)}U^{(\pm)}$ 
act as projection operators and we find a nice geometric interpretation in terms of $f$-structures.

\subsection{A Yano $f$-structure}

The fact that the matrices $U^{(\pm)}$ (and $V^{(\pm)}$) are 
degenerate and satisfy (\ref{sumer1}) and (\ref{sumer1c}),
\ber\nonumber
&&U^{(+)}V^{(+)}=-\textrm{diag}(1,0,1,1), \quad V^{(+)}U^{(+)}=-\textrm{diag}(0,1,1,1),\\[1mm]
&&U^{(-)}V^{(-)}=-\textrm{diag}(1,1,1,0), \quad V^{(-)}U^{(-)}=-\textrm{diag}(1,1,0,1)
\eer{info}
prevents a direct interpretation in terms of complex structures on 
the tangent space $TM$.
We are led to consider endomorphisms on $TM\oplus TM$ 
and the weaker $f$-structures instead. 
The following $8d\times 8d$ matrices are $f$-structures in the sense of Yano
\cite{Yano:1961}:
\beq
{\cal F}_{(\pm)}:=\left(\begin {array}{cc}
0&U^{(\pm)}\cr
V^{(\pm)}&0\end{array}\right)\quad 
\implies {\cal F}^{3}_{(\pm)}+{\cal F}_{(\pm)}=0~.
\eeq{fst}
This follows directly from conditions in (\ref{sumer1}).
Moreover, $-{\cal F}^2_{(\pm)}$ and $1+{\cal F}^2_{(\pm)}$ 
define integrable distributions, as can be shown using (\ref{sumer1}) and \re{Nije}.
More explicitly: Using the projectors \re{projectors}, 
the conditions \re{info} may be written as 
\beq
\hat P_{\pm} = 1 + V^{(\pm)} U^{(\pm)} ~,\quad
P_{\pm} = 1 + U^{(\pm)}V^{(\pm)} ~.
\eeq{projs}
Then we may define
\begin{equation}
 m_{(\pm)}:=1+{\cal F}^{2}_{(\pm)} = \left(
 \begin{array}{cc}
 P_\pm & 0 \\ 0 & \hat P_\pm
 \end{array}\right), \quad
 l_{(\pm)}:=-{\cal F}^{2}_{(\pm)} = \left(
 \begin{array}{cc}
 1-P_\pm & 0 \\ 0 & 1-\hat P_\pm
 \end{array}\right).
\end{equation}
These fulfill
\beq
 l_{(\pm)} + m_{(\pm)} = 1,\quad
 l^2_{(\pm)} = l_{(\pm)}, \quad m^2_{(\pm)} = 
 m_{(\pm)}~,\quad l_{(\pm)} m_{(\pm)} =0
\eeq{fund1}
and
\beq
 {\cal F}_{(\pm)} l_{(\pm)} = l_{(\pm)} {\cal F}_{(\pm)} = 
 {\cal F}_{(\pm)}, \quad m_{(\pm)}{\cal F}_{(\pm)} =
 {\cal F}_{(\pm)} m_{(\pm)} =0.
\eeq{fund2}
The operators $l_{(\pm)}$ and $m_{(\pm)}$ applied to the tangent space
at each point of the manifold are complementary projection operators
and define complementary distributions in the sense of Yano:
$\Lambda_\pm$, the first fundamental distribution, and $\Sigma_\pm$,
the second fundametal distribution, corresponding to $l_\pm$ and
$m_\pm$, of dimensions $6d$ and $2d$, respectively.

Let $\mathcal{N}_{{\cal F}_{(\pm)}}$ denote the Nijenhuis tensor for
the $f$-structures ${\cal F}_{(\pm)}$. By a theorem of Ishihara and
Yano \cite{IshiharaYano} we have that
\begin{enumerate}
 \item[i.]
 $\Lambda_\pm$ is integrable iff $m^i_{(\pm)l} 
 \mathcal{N}_{\mathcal{F}_{(\pm)} jk}^l=0$,
 \item[ii.]
 $\Sigma_\pm$ is integrable iff 
 $\mathcal{N}_{\mathcal{F}_{(\pm)} jk}^i m^j_{(\pm)l} m^k_{(\pm) m} =0$.
\end{enumerate}
From the definition of the $f$-structures in \re{fst}, one can 
derive that these two conditions are fulfilled. Hence, the 
distributions $\Lambda_\pm$ and $\Sigma_\pm$ are integrable. 

\subsection{Additional twisted supersymmetry}
In a previous paper \cite{Goteman:2009xb} we investigated the special case of four-dimensional target space and required the transformations (\ref{2}) to be linear. There, it was found that no solution with interesting geometry exists which possesses additional supersymmetry. On the other hand, one could impose additional twisted linear supersymmetry $[\delta, \delta]= -\partial$ for a solution with interesting geometrical properties. 

In the general case treated in this paper, we have found that additional supersymmetry can indeed be imposed. But also additional twisted supersymmetry could be considered. The difference would be that contraint (\ref{bub1_noncov}) would receive a minus sign,
\ber
\mathcal{M}(U^{(\pm)},V^{(\pm)})^i_{jk} 
\bar{\mathbb{D}}_{\pm}\mathbb{X}^j \mathbb{D}_{\pm}\mathbb{X}^k \nonumber 
&=&\phantom{+}\left[\left(UV\right)^{(\pm)i}_{~j} - \delta^i_j
\right]\bbDB{\pm}\bbD{\pm}\bbX{j} \nonumber\\ 
&& +\left[\left(VU\right)^{(\pm)i}_{~j} - \delta^i_j\right]\bbD{\pm}\bbDB{\pm}\bbX{j}~.
\eer{constraint3_pseudosusy}
with the effect that the structure defined in (\ref{fst}) would be a $f$-structure of hyperbolic type,
\beq
	\mathcal{F}_{(\pm)}(\mathcal{F}_{(\pm)}^2 -1)=0,
\eeq{f_hyperbolic}
that is, generalizations of product structures instead of complex structures.

\section{Invariance of the action}
\label{inv}

The bihermitean geometry of \cite{Gates:1984nk} is derived from the
$(1,1)$ sigma model via two requirements: closure of the algebra and
invariance of the action. More precisely, the supersymmetry algebra implies 
the existence of the complex structures, whereas invariance 
the action implies the bihermiticity of the metric and the covariant constancy 
of the complex structures. Similarly, for $(4,4)$ supersymmetry,
the algebra implies that the transformations are given in terms 
of left and right hypercomplex structures whereas invariance of
the action implies the metric is hermitean with respect to all of
these structures and the left and right connections preserve the
the left and right structures respectively. When, in
later sections, we use the knowledge from \cite{Gates:1984nk} in
understanding our algebra conditions on-shell we can thus use the
existence of a hypercomplex structures freely, but only
require them to be covariantly constant if we assume that the action
is invariant.

At the manifest $(2,2)$ level the discussion of additional
supersymmetries in the  model  with (anti)chiral fields (the
hyperk\"ahler case) follows similar lines \cite{Hull:1985pq}. Extra
supersymmetries lead to new complex structures as part of the
conditions for closure of the algebra and invariance of the $(2,2)$
action leads to to the requirement that they are covariantly constant 
and that the metric is bihermitean. \bigskip

When the complex structures commute and the sigma model is describable
in $(2,2)$ superspace using (an equal number of) chiral and twisted
chiral superfields, $(4,4)$ supersymmetry comes at the price of extra
conditions on the potential $K$ \cite{Gates:1984nk}. This is also true
for the linear-transformation model in \cite{Goteman:2009xb}. We 
expect the same to be true here.

The action \re{act1} is invariant under the supersymmetry 
transformations (\ref{MNdef}) provided that 
\begin{equation}
\label{hermiticity}
\left(K_{i}U^{(+)i}{}_{[j}\right){}_{k]}=0, \quad j,k \neq a,
\end{equation}
and analogously for $U^{(-)}$ and $V^{(\pm)}$. We can write this out as
(\ref{2}) a system of equations for $K$:
\beq
	\bigr(K_{a}f^i_{[j} + K_{a'} \tilde{g}^{a'}_{[j} + K_{\bar a'}\bar{\tilde h}^{\bar a'}_{[j}\bigr){}_{k]} = 0
\eeq{inv_simple}
plus analogous relations from $U^{(-)}$ and $V^{(\pm)}$.

The conditions \re{inv_simple} (or \re{hermiticity}) have to be satisfied
for the generalized K\"ahler potential $K$ to allow $(4,4)$
supersymmetry in a model with noncommuting complex structures whose
commutator has empty kernel. In this sense it plays a similar role to
the Monge-Amp\`ere equation for models with vanishing torsion.

In the four-dimensional case with linear twisted supersymmetry transformations, it turned
out to be possible to solve \re{inv_simple}, (see \cite{Goteman:2009xb})
but this is much harder in general. However, when the curl of $\tilde
g$ and $\tilde h$ vanish, the condition has an interpretation on
$TM\oplus TM$ much like a hermiticity condition, which we now turn to.

We combine the Hessian $K_{ij}$ of the K\"ahler potential into an
antisymmetric tensor on $ \mathfrak{B}$ on $TM\oplus TM$ as 

\beq
 \mathfrak{B} = \left(\begin{array}{cc}
0 & K \\ -K^t & 0
\end{array}\right).
\eeq{omega}
The relation (\ref{hermiticity}) can be used to show that off-shell 
the $f$-structures \re{fst} preserve $ \mathfrak{B}$ on a subspace 
projected out by the second fundamental projection operators 
$l_{(\pm)}$ defined in \re{fund2},
\beq
l_{(\pm)} \, \mathcal{F}^t_{(\pm)} \mathfrak{B} 
\mathcal{F}_{(\pm)} \, l_{(\pm)} = l_{(\pm)} \, \mathfrak{B} \, l_{(\pm)}.
\eeq{preserve}
This may be easily verified using  \re{sumer1}, which implies
$V^t K^t U = - K$ (except for one column and one row). 

\section{On-shell interpretation of the algebra constraints}
 \label{OnShell}

In this section we discuss two main issues: How the conditions derived
in section \ref{algcstrts} have a larger set of solutions on-shell,
and the relation to the underlying (hermitean) bi-hypercomplex
geometry derived in \cite{Gates:1984nk}. In spirit the treatment is
similar to both the $(1,1)$ discussion in \cite{Gates:1984nk} of
extended supersymmetry  and to the hyperk\"ahler derivation in
\cite{Hull:1985pq}: In   \cite{Gates:1984nk}  it was found that the
left and right complex structures had to commute to get off-shell
closure since the algebra gives a term proportional to this commutator
times the field-equations. In \cite{Hull:1985pq} it was found that
field equations as well as conditions from the invariance of the
action were needed for closure of the algebra of non-manifest
additional supersymmetries.

Below we separate the conclusions we may draw from closure of the
algebra only and those where in addition we need invariance of the
action.

\subsection{On-shell algebra}
\label{onprel}
In this subsection we use a coordinate transformation to derive an 
explicit relation between the components of the transformation 
matrices and the underlying hypercomplex structure.
The field equations that follow from the action \re{act1} are
\beq
\bbDB{+}K_a=0~,~~\bbD{+} K_{\bar a}=0~,~~\bbDB{-} K_{a'}=0~,~~\bbD{-}K_{\bar a'}=0~.
\eeq{eoms}
These imply that on-shell, $K_a$ is a semichiral superfield on equal
footing with $\bbX{a}$; we may change coordinates to a
left-holomorphic or right-holomorphic basis with coordinates
$Z^A=\{\bbX{a}, Y_a:=K_a\}$ or $Z^{A'}=\{\bbX{a'}, Y_{a'}:=K_{a'}\}$,
respectively \cite{Bogaerts:1999jc}. In the left basis, the
$\delta^+,\bar\delta^+$ transformations become very simple, whereas in
the right basis, the $\delta^-,\bar\delta^-$ transformations become
simple. Since $K(\bbX{a},\bbX{a'})$ is the generating function for the
transformation between the bases, on-shell it is sufficient to study
the transformations that are simple in one particular basis.

The ansatz for the $\delta^+,\bar\delta^+$ transformations is simple
in the left basis:
\beq
\delta^+ Z^A=0~,~~\delta^+ \bar Z^{\bar A}=\epsilon^+\bbD{+}\fbb^{\bar A}~,~~
\bar\delta^+ Z^A=\bar\epsilon^+\bbDB{+}\fb^{A}~,~~\bar\delta^+ \bar Z^{\bar A}=0~.
\eeq{leftansatz}
Closure of this part of the algebra is very simple; it implies
\beq
\fb^A{}_{\bar B}\fbb^{\bar B}{}_{C}=-\delta^A_C~,~~
\eeq{almostcomplex}
and 
\beq
\fb^A{}_{C[\bar B }\fb^C{}_{\bar D]} = 0~,~~
\eeq{nijen}
where $\fb^A{}_{\bar B}$ again denotes derivation with respect to $\bar Z^{\bar B}$.
These are precisely the conditions found in section 10 of \cite{Hull:1985pq}, and imply that
\beq
 J_{(+)}^{(1)} =
 \left( \begin{matrix} 
 0& \fb^A{}_{\bar B} \\
 \fbb^{\bar A}{}_{B} &0 \\
 \end{matrix}\right)~,~~
 J_{(+)}^{(2)}= \left( \begin{matrix} 
 0& i\fb^A{}_{\bar B} \\
 -i\fbb^{\bar A}{}_{B} &0 \\
 \end{matrix}\right)~,~~
 J_{(+)}^{(3)}= \left( \begin{matrix} 
 i{\mathbb I}& 0 \\
 0&- i{\mathbb I} \\
 \end{matrix}\right)
\eeq{hypercom}
generate an integrable hypercomplex structure.
Similarly, in the right basis, the $\delta^-$ transformations generate a second
integrable hypercomplex structure so that in total we get a bi-hypercomplex structure, 
\beq
J_{(\pm)}^{(A)}J_{(\pm)}^{(B)}=-\delta^{AB}+\epsilon^{ABC}J_{(\pm)}^{(C)}~.
\eeq{sut}

We still need to impose the $[\delta^+,\delta^-]$ part of the algebra and
want to compare to the off-shell transformations (\ref{2}). For both of these tasks,
we need to go back to the $\bbX{a},\bbX{a'}$ coordinate basis. 
For illustrative purposes,
we focus on $\bar\delta^+$. Comparing (\ref{2}) and (\ref{leftansatz}), 
we immediately find
that on-shell
\beq
f^a(\bbX{i})=\fb^a(\bbX{a},\bbXB{\bar a},K_a(\bbX{i}),K_{\bar a}(\bbX{i}))~.
\eeq{fisfb}
Off-shell, $f^a$ may differ from $\fb^a$ by a factor $\Delta f^a$, 
which satisfies $\bbDB+(\Delta f^a (\bbX{a},K_a(\bbX{i}))) =0$ on-shell. 
This gives an off-shell ambiguity in $f^a$. We also have (trivially) that 
$\bar\delta^+\bbXB{\bar a}=0$. 
Next we have
\beq
\bar\delta^+\bar Y_{\bar a} = K_{\bar a b}\bar\delta^+\bbX{b}+
K_{\bar a R}\bar\delta^+\bbX{R}=0
\eeq{dby}
and 
\beq
\bar\delta^+Y_a := K_{a b}\bar\delta^+\bbX{b}+K_{a R}
\bar\delta^+\bbX{R}=\bar\epsilon^+\bbDB+f_a
= \bar\epsilon^+(f_{a\bar b}\bbDB+\bbXB{\bar b}+f_{aR}
\bbDB+\bbX{R})~,
\eeq{dy}
where $f_a(\bbX{i}):=\fb_a(\bbX{a},\bbXB{\bar a},
K_a(\bbX{i}),K_{\bar a}(\bbX{i}))$.
We can rewrite these equations as
\beq
K_{LR}\bar\delta^+\bbX{R}=
 \bar\epsilon^+ \left( \begin{matrix} 
 f_{a\bar b}\bbDB+\bbXB{\bar b}+f_{aR}\bbDB+\bbX{R}-
 K_{ab}\bbDB+f^b \\
 -K_{\bar a b}\bbDB+f^b
 \end{matrix}\right)~,
\eeq{dbxlr}
where the matrix $K_{LR}$ is defined as in (\ref{KLR}).
Since $K_{LR}$ is invertible, we can find the on-shell
transformations $\bar\delta^+\bbX{R}$. To find the corresponding 
functions $\tilde g^{a'}_{b'}$ and
$\bar{\tilde{h}}^{\bar a'}_{\bar b'}$ in (\ref{2}), since we are on-shell, 
we need to eliminate one type of term, \eg $\bbDB+\bbXB{\bar b}$, 
using the field equations.\footnote{We assume that $K_{a\bar b}$ is invertible, 
otherwise, we would need to eliminate another type of term, but the net effect 
would be the same.} Then (\ref{dbxlr}) becomes 
\beq
K_{LR}\bar\delta^+\bbX{R}=
 \bar\epsilon^+ \left( \begin{matrix} 
 -(f_{a\bar c}-K_{ab}f^b_{\bar c})(K^{-1})^{\bar c d}K_{dR} 
 +f_{aR}-K_{ab}f^b_R \\
K_{\bar a b}f^b_{\bar c}(K^{-1})^{\bar c d}K_{dR} -K_{\bar a b}f^b_R
 \end{matrix}\right)\bbDB+\bbX{R}~~.
\eeq{dbxr}
and we find
\ber
\tilde g^{e'}_{f'}&=&(K^{-1})^{e'a}[f_{af'}-K_{ab}f^b_{f'} 
-(f_{a\bar c}-K_{ab}f^b_{\bar c})(K^{-1})^{\bar c d}K_{df'}]\nonumber\\
&&-(K^{-1})^{e'\bar a}[K_{\bar ab}f^b_{f'}
-K_{\bar a b}f^b_{\bar c}(K^{-1})^{\bar c d}K_{df'}]~,\nonumber\\
\bar{\tilde{h}}^{\bar e'}_{\bar f'}&=&(K^{-1})^{\bar e'a}[f_{a\bar f'}-K_{ab}f^b_{\bar f'} 
-(f_{a\bar c}-K_{ab}f^b_{\bar c})(K^{-1})^{\bar c d}K_{d\bar f'}]\nonumber\\
&&-(K^{-1})^{\bar e'\bar a}[K_{\bar ab}f^b_{\bar f'}
-K_{\bar a b}f^b_{\bar c}(K^{-1})^{\bar c d}K_{d\bar f'}]~,
\eer{ghonshell}
as well as the constraints
\ber
0&=&(K^{-1})^{e'a}[f_{a\bar f'}-K_{ab}f^b_{\bar f'} 
-(f_{a\bar c}-K_{ab}f^b_{\bar c})(K^{-1})^{\bar c d}K_{d\bar f'}]\nonumber\\
&&-(K^{-1})^{e'\bar a}[K_{\bar ab}f^b_{\bar f'}
-K_{\bar a b}f^b_{\bar c}(K^{-1})^{\bar c d}K_{d\bar f'}]~,\nonumber\\
0&=&(K^{-1})^{\bar e'a}[f_{af'}-K_{ab}f^b_{f'} 
-(f_{a\bar c}-K_{ab}f^b_{\bar c})(K^{-1})^{\bar c d}K_{df'}]\nonumber\\
&&-(K^{-1})^{\bar e'\bar a}[K_{\bar ab}f^b_{f'}
-K_{\bar a b}f^b_{\bar c}(K^{-1})^{\bar c d}K_{df'}]~.
\eer{constraintfromonshell}
In a similar way, we can find $g^a_b, h^a_b$ as well as their complex conjugates.
The full set of relations will now be discussed in the original coordinates $\bbX{i}$.
\subsection{Closure modulo field-equations and relations from invariance of the action}
Though conceptually simple, the final expressions that we found
(\ref{ghonshell})--(\ref{constraintfromonshell}) are rather involved
and complicate the discussion on the on-shell $[\delta^+,\delta^-]$
algebra. Here we present an alternative description that uses only
$\bbX{i}$ coordinates and relates directly to the bi-hypercomplex
geometry of \cite{Gates:1984nk}. We start from the ansatz \re{2} and
only use the field equations to show that the conditions from closure
of the algebra have more solutions on-shell. Whereas in the previous
subsection discussing the on-shell algebra, it was convenient to change
coordinates, here it turns out to be convenient to change the basis
for the covariant derivatives.

Recall the field equations (\ref{eoms})
\ber
K_{ai}\bbDB{+}\bbX{i}=0~,
&& K_{a'i}\bbDB{-}\bbX{\bar i}=0,\cr
K_{\bar a i}\bbD{+}\bbX{i}=0~,
&& K_{\bar a' i}\bbD{-}\bbX{i}=0~.
\eer{eom}

These equations are first order in spinorial derivatives. To be able
to use them to understand the conditions \re{list1} 
\re{bub1_noncov}, which contain second order spinorial deivatives, we
must differentiate \re{eom}. We are then faced with the task of
relating the plus/minus connections to second and third derivatives of
the generalized K\"ahler potential $K$. Since the metric is a
nonlinear function of the Hessian of $K$, this is not easy. Instead we
choose to express the on-shell condition in terms of the complex
structures $J_{(\pm)}$ defined in (\ref{jpm}) and use
$\nabla^{(\pm)}J_{(\pm)}=0$ to relate them to the connections
(assuming invariance of the action).

We introduce a real basis for the spinor derivatives:

\beq
\bbD{\pm}:=\half(D_\pm-iQ_\pm)~,
\eeq{bchg}
then \re{eom} becomes\footnote{Note however that we use full 
$(2,2)$ superfield expressions in, \eg \re{Sdef}; we can reduce to $(1,1)$ superspace
by restricting to superfields to depend only on half the spinor coordinates.} 
\ber
Q_+\bbX{R}&=&J_{(+)k}^RD_+\bbX{k}\cr
Q_-\bbX{L}&=&J_{(-)k}^LD_-\bbX{k}
\eer{Sdef}
where we have introduced (components of) the complex structures 
$J_{(\pm)}$ as defined in section \ref{prelim}.

The semichiral conditions rewritten in terms of the real operators 
\re{bchg} and \re{jjj} read 
\ber
Q_+\bbX{L}&=&jD_+\bbX{L}\cr
Q_-\bbX{R}&=&jD_-\bbX{R}~.
\eer{semiagain}
Combining this with \re{Sdef} and \re{jpm} we find that on-shell
\beq
Q_\pm\bbX{}:=Q_\pm\left(\begin{array}{c} \bbX{L}\cr \bbX{R}
\end{array}\right)=J_{(\pm)}D_\pm\left(\begin{array}{c} 
\bbX{L}\cr \bbX{R}\end{array}\right)=J_{(\pm)}D_\pm\bbX{}~,
\eeq{deru}
which using \re{bchg} implies
\ber\nonumber
\bbD{\pm}\bbX{i}&=&\bar \pi^{(\pm)i}_{k}D_\pm \bbX{k}\\[1mm]
\bbDB{\pm}\bbX{i}&=&\pi^{(\pm)i}_{k}D_\pm \bbX{k}
\eer{proj}
where we have introduced the projection operator
\beq
\pi:=\half \left( \one + iJ\right)~,
\eeq{projo}
and its complex conjugate.
\subsection{Relations to bi-hypercomplex geometry}
\label{bihyp}

In subsection \ref{onprel} we constructed the bi-hypercomplex
structures directly in terms of the transformations of the left and
right holomorphic coordinates, and related bi-hypercomplex structures
to the $f$-structures implicitly by constructing the tensors in the
ansatz \re{2} in terms of the same transformations. In this subsection
we analyze the relation using the real basis; this makes some aspects
clearer while complicating others.

From the $N=(1,1)$ analysis of \cite{Gates:1984nk} we know that when
the model has $(4,4)$ supersymmetry there exists an $SU(2)$
worth of left and right
complex structures $(J_{(\pm)}^{(1)},J_{(\pm)}^{(2)},J_{(\pm)}^{(3)})$
on the $4d$ dimensional space, satisfying the bi-hypercomplex algebra
(\ref{sut}). We now relate the $f$-structures to $J_{(\pm)}^{(A)}$.

The complex structures $J_{(\pm)}$ are part of the $SU(2)$ worth of
complex structures, and we set $J_{(\pm)}^{(3)}:=J_{(\pm)}$. In the
real basis \re{bchg}, the additional supersymmetries take the form
\beq
\delta_s\bbX{}:= \delta^{\pm}\bbX{}+\bar\delta^{\pm}\bbX{}=
\frac12 \left[\left(J^{(1)}_{(\pm)}+iJ^{(2)}_{(\pm)}\right)
\epsilon^\pm D_\pm\bbX{}+\left(J^{(1)}_{(\pm)}-
iJ^{(2)}_{(\pm)}\right) \bar\epsilon^{\pm}D_\pm\bbX{}\right],
\eeq{add}
Identifying (\ref{add}) with (\ref{MNdef}) we deduce that
\ber
\frac 1 2\left(J^{(1)}_{(\pm)}-iJ^{(2)}_{(\pm)}\right)&=&
U^{(\pm)}\pi^{(\pm)}~,\nonumber\\[3mm]
\frac 1 2\left(J^{(1)}_{(\pm)}+iJ^{(2)}_{(\pm)}\right)&=&
V^{(\pm)}\bar\pi^{(\pm)}~.
\eer{cmp}
This relation implies
\ber
(UV)^{(\pm)}\bar\pi^{(\pm)}&=&-\bar\pi^{(\pm)}\cr
(VU)^{(\pm)}\pi^{(\pm)}&=&-\pi^{(\pm)}~.
\eer{pisky}
A further consequence of the algebra \re{sut} is, \eg that
\ber
U^{(\pm)}\pi^{(\pm)}=\bar\pi^{(\pm)}U^{(\pm)}\pi^{(\pm)}~,
\quad V^{(\pm)}\bar\pi^{(\pm)}=\pi^{(\pm)}V^{(\pm)}\bar\pi^{(\pm)}~.
\eer{bbb}
On $TM\oplus TM$ we have that 
\ber
\half\left(\begin{array}{cc}0&J^{(1)}_{(\pm)}-iJ^{(2)}_{(\pm)}\cr
J^{(1)}_{(\pm)}+iJ^{(2)}_{(\pm)}&0\end{array}\right)={\cal F}_{(\pm)}
\left(\begin{array}{cc}\bar\pi^{(\pm)}&0\cr
0&\pi^{(\pm)}\end{array}\right)=:{\cal F}_{(\pm)}\Pi_{(\pm)}~,
\eer{tot}
and the relations \re{bbb} can be used to show that both sides square 
to -$\Pi_{(\pm)}$.

Finally, assuming that the action is invariant we have  
$\nabla^{(\pm)} J^{(A)}_{(\pm)}=0$, (see \re{cs}) which implies that
\ber
&&\nabla^{(\pm)}U^{(\pm)}\pi^{(\pm)}=0\cr
&&\nabla^{(\pm)}V^{(\pm)}\bar\pi^{(\pm)}=0~,
\eer{nablaplus}

The equations \re{cmp}--\re{tot} expresses the relation between the
bi-hypercomplex geometry and the extra supersymmetries \re{2}. The
relation does not seem to be one-to-one since only, \eg
$U^{(+)}\pi^{(+)}$ enters. However, the particular form \re{MNdef2} of
$U^{(\pm)}$ may be used in combination with the explicit expressions
\re{jpm} of $J_{(\pm)}$ to show that all of $U^{(\pm)}$ is in fact
determined by $J^{(A)}_{(\pm)}$. This is evident from the explicit
expressions for the components of $U^{(+)}$ in section \ref{onprel}.

\subsection{On-shell interpretation of the constraints.}
\label{onshell}
On-shell, there are more cases when the algebra of the extra supersymmetries close, 
in analogy to, \eg models written in terms of (anti)chiral fields.
To illustrate the line of argument we first discuss \re{bub1_noncov}.  

Modulo the curl-part,
\beq
U^{(\pm)l}_{~j}V^{(\pm)i}_{[k,l]}-V^{(\pm)l}_kU^{(\pm)j}_{[j.l]}
\eeq{curl} 
we may use  \re{sumerien} to rewrite \re{bub1_noncov}  as
\ber
&-&\left[\left(VU\right)^{(\pm)i}_{~j,k}-\left(UV\right)^{(\pm)i}_{~k,j}\right]
\bar{\mathbb{D}}_{\pm}\mathbb{X}^j \mathbb{D}_{\pm}\mathbb{X}^k\\[1mm]
&+&\left[\left(UV\right)^{(\pm)i}_{~j}+\delta^i_j\right]
\bbDB{+}\bbD{\pm}\bbX{j}+\left[\left(VU\right)^{(\pm)i}_{~j}+
\delta^i_j\right]\bbD{\pm}\bbDB{+}\bbX{j}=0.
\eer{rewrite}
Since the LHS is
\ber
\bbDB{\pm}\left[\left(UV\right)^{(\pm)i}_{~j}\bbD{\pm}\bbX{j}\right]+
\bbD{\pm}\left[\left(VU\right)^{(\pm)i}_{~j}\bbDB{\pm}\bbX{j}\right]+
\{\bbDB{\pm},\bbD{\pm}\}\bbX{i}
\eer{LHS}
and we know from \re{proj} and \re{pisky} that on-shell the square 
brackets become $-\bbD{\pm}\bbX{i}$ and $-\bbDB{\pm}\bbX{i}$ 
respectively, we see that the LHS vanishes on-shell. It remains to 
consider the terms in \re{curl}. 

Writing the term out in full, including the derivatives, we have
\ber\nonumber
&&(U^{(\pm)l}_{~j}V^{(\pm)i}_{[k,l]}-V^{(\pm)l}_kU^{(\pm)i}_{[j,l]})
\bbDB{\pm}\bbX{j}\bbD{\pm}\bbX{k}\\[1mm]
&=& (U^{(\pm)l}_{~j}V^{(\pm)i}_{[k,l]}-V^{(\pm)l}_kU^{(\pm)i}_{[j,l]})
\pi^{(\pm)j}_p\bar\pi^{(\pm)k}_q D_{\pm}\bbX{p}D_{\pm}\bbX{q}
\eer{curlon}
Using the relations \re{cmp} and \re{bbb} it is possible to show that one 
can replace all the $U$'s and $V$'s by, e.g., combinations of $\pi^{(\pm)}$'s 
and $J_{(\pm)}^{(1)}$ yielding the following expression for the curl-terms:
\beq
\left(J_{k}^{(1)i}{\mathcal N}(\bar\pi)^k_{rq}J_{j}^{(1)r}
\pi^{j}_p-J_{k}^{(1)i}{\mathcal N}(\pi)^k_{rp}J_{j}^{(1)r}\bar\pi^{j}_q
+{\mathcal N}(J^{(1)})^i_{jk}\pi^{j}_p\bar\pi^{k}_q\right)
D \bbX{p}D \bbX{q}~,
\eeq{curlon2}
where the $(\pm)$-indices were omitted for clarity. The integrability
of the $J_{(\pm)}^{(A)}$'s means that all the Nijenhuis-tensors and
thus all of terms in \re{curlon2} vanish. We thus see that on-shell
\re{bub1_noncov} implies no new constraints.

Next we consider \re{list1}. Off-shell we had to set the terms with
independent structures separately to zero \re{sum}. On-shell we find
no  conditions  on the tensors if we also assume invariance of the
action.

The RHS of \re{list1} is

\beq
[U^{(+)},U^{(-)}]^{i}_j\bar{\bbnab}_+^{(-)}\bbDB{-}\bbX{j}~,
\eeq{good}
where ${\bbnab}^{(-)}_\pm$ is the pull-back of the minus-covariant
derivative $\nabla^{(-)}_{i}$ in the $\bbD{\pm}$ basis. We want to
avoid the off-shell conclusion that the commutator vanishes and
observe that the commutator multiplies something that looks like a
field equation. However, we have to use \re{deru} to see if it
actually vanishes on-shell.

In the remainder of this section, we use the conditions that follow
from invariance of the action \cite{Gates:1984nk}, which imply that
the metric is hermitean with respect to all the complex structures and
the connections $\Gamma^{(\pm)}$ preserve the hypercomplex
structures\footnote{This is equivalent to restricting the holonomy of
the connections $\Gamma^{(\pm)}$ to a symplectic group.} $J_{(\pm)}$:
$\nabla^{(\pm)} J_{(\pm)} =0$. A straightforward calculation shows
that\footnote{Here the operator $\nabla^{(-)}_\pm$ is the pullback in
the $D_\pm$-basis.}
\beq
\bar{\bbnab}_+^{(-)}\bbDB{-}\bbX{i}=-\half\left\{\pi^{(-)},\pi^{(+)}
\right\}^i_k\nabla_+^{(-)}D_-\bbX{k}~.
\eeq{fin}
To lowest order, the RHS is proportional to the $(1,1)$ field equation. 
Since it is written in manifest $(2,2)$ form, one may expect that it also vanishes
to all orders. In fact, the $(2,2)$ relation
\beq
\{Q_+,Q_-\}\bbX{i}=0~,
\eeq{final}
has the on-shell content
\beq
[J_{(-)},J_{(+)}]^i_j\nabla^{(-)}_+D_-\bbX{j}=0~,
\eeq{finale}
where again covariant constancy of the complex structures is used.
Since the commutator is invertible in a model with only semichiral fields,
\beq
\nabla^{(-)}_+D_-\bbX{j}=0~,
\eeq{FE}
and that the RHS of \re{fin} vanishes. 

Using the connections with skew torsion $T=\pm\half dB$  
we have from the definition \re{plusminuscon} that the LHS of \re{list1} is
\ber\nonumber
&&\widehat{\mathcal{M}}(U^{(+)},U^{(-)})^i_{jk}
\bbDB{+}\bbX{j} \bbDB{-}\bbX{k}=\\[1mm]\nonumber
 &&\left(U^{(+)l}_j \hcd{l}^{(-)} U^{(-)i}_k - U^{(-)l}_k 
 \hcd{l}^{(+)} U^{(+)i}_{j} 
 - U^{(+)i}_l \hcd{j}^{(-)} U^{(-)l}_{k} 
 + U^{(-)i}_l \hcd{k}^{(+)} U^{(+)l}_{j}\right)\bbDB{+}\bbX{j} 
 \bbDB{-}\bbX{k}~.\\
 &&
\eer{mhat}
Given the results for the RHS, the appropriate projections of 
$\widehat{\mathcal{M}}(U^{(+)},U^{(-)})$ thus have to vanish. 
However, we know from \re{proj} that on-shell 
\beq
\bbDB{+}\bbX{j} \bbDB{-}\bbX{k}=\pi^{(+)j}_{l}D_+\bbX{l}\pi^{(-)k}_{s}D_-\bbX{s}~,
\eeq{rep}
and invoking invariance of the action, we may use \re{nablaplus} 
to conclude that then indeed $\widehat{\mathcal{M}}(U^{(+)},U^{(-)})=0$.

In summary our result is very similar to the hyperk\"ahler 
discussion in \cite{Hull:1985pq}, we need to invoke invariance of the 
action to show that there are more solutions on-shell to the conditions 
from the algebra\footnote{In hyperk\"ahler case the the algebra {\em only} closes on-shell.}.

The only constraints we get on the transformation matrices on-shell 
for invariant actions are the integrability condition 
\beq
\mathcal{N}(U^{(\pm)})^i_{jk} \pi^{(\pm)j}_l \pi^{(\pm)k}_m 
D_{\pm}\mathbb{X}^l D_{\pm}\mathbb{X}^m = 0~.
 \eeq{leftrightonshell}
together with the identification (\ref{cmp}).

\section{Discussion}
\label{disk}

Throughout this paper, the arbitrary entries in the transformation 
matrices $U^{(\pm)}$ (and $V^{(\pm)}$) were set to zero. Off-shell, 
this has the advantages of yielding geometric structures on the full target-space. 
Keeping the arbitrariness would restrict the features (\eg integrability) 
of these structures to certain subspaces.

We have identified new geometric structures on the target-space of 
sigma models written in terms of semichiral fields.
These structures arise when we study additional off-shell supersymmetries. 
We have discussed the $f$-structures as living on the sum of two copies 
of the tangent bundle $TM\oplus TM$. Clearly one would like to identify 
the relation to generalized complex geometry on $TM\oplus T^*\! M$. 
Formally, this may be achieved using the existence of a metric \cite{Lindstrom:2005zr}
\beq
g= \Omega [J_{(+)},J_{(-)}]~,
\eeq{met}
where 
\beq
\Omega :=\left(\begin{array}{cc}0 &2iK_{LR}\cr 
-2iK_{RL}&0\end{array}\right)~.
\eeq{GO}
We use $g$ to relate $TM$ and $T^*\! M$ to write 
$\cal F$ as an $f$-structure on $TM\oplus T^*\! M$:
\beq
\tilde{\cal F} :=\left(\begin{array}{cc}0 &Ug^{-1}\cr 
gV&0\end{array}\right)~.
\eeq{GCF}
We plan to return to the geometry of $f$-structures in the context of
generalized complex geometry in a later publication.\bigskip

A related question concerns the condition for invariance of the
action. As we have shown for a subclass of our transformations, this
amounts to the conservation of an antisymmetric tensor $\mathfrak{B}$
on certain subspaces of $TM\oplus TM$ by the $f$-structures. Again,
the corresponding object on $TM\oplus T^*\! M$ can be found using the
metric $g$:
\beq
 \tilde{\mathfrak{B}} = \left(\begin{array}{cc}
0 & Kg^{-1} \\ -gK^t & 0
\end{array}\right).
\eeq{Eagain}
It remains to clarify where this object fits into the generalized
complex picture. This also ties in with the question of how the
conditions for invariance that we have described relate to those found
in \cite{Lindstrom:1994mw}, where $(4,4)$ models with auxiliary fields
are discussed. \bigskip

In the precursor to this article \cite{Goteman:2009xb} where the
nonmanifest transformations were linear and the target space was four-dimensional, there was no interesting solution with additional supersymmetry. Additional twisted supersymmetry could be imposed, however. The target-space was then seen to carry indefinite signature metric and 
vanishing three form $H$, the geometry being pseudo-hyperk\"ahler. 
In the present paper, where the target space is $4d$-dimensional, the transformations close to an ordinary supersymmetry algebra if $d>1$, i.e. the dimension of the target space is larger than four. This stems from the fact that a complex number $a$ can never fulfill $a \bar a=-1$, whereas for a matrix $A$ with complex conjugated components $\bar A$, this could indeed be fulfilled. 
We could also have considered a twisted supersymmetry in the general case. The result would have been hyperbolic $f$-structures, a generalization of the result in \cite{Goteman:2009xb}.

\bigskip

{\bf Acknowledgements}: We thank Rikard von Unge and Maxime Zabzine
for discussions. The research of UL was supported by VR grant
621-2006-3365. The research of MR and IR was supported in part by NSF
grant no. PHY-06-53342. UL, MR, IR are happy to thank the 2008 and
2009 Simons Summer Workshops for creating an environment that
stimulated some of this research.


\bigskip

\begin{thebibliography}{9}
\newcommand{\np}{{\em Nucl.\ Phys.\ }}
\newcommand{\pr}{{\em Phys.\ Rev.\ }}
\newcommand{\cmp}{{\em Commun.\ Math.\ Phys.\ }}
\newcommand{\pl}{{\em Phys.\ Lett.\ }}

\bibitem{Gates:1984nk}
 S.~J.~Gates, C.~M.~Hull and M.~Ro\v cek,
 {\em Twisted Multiplets And New Supersymmetric Nonlinear Sigma Models},
 Nucl.\ Phys.\ B {\bf 248}, 157 (1984).
 
\bibitem{Howe:1985pm}
 P.~S.~Howe and G.~Sierra,
 {\em Two-Dimensional Supersymmetric Nonlinear Sigma Models With Torsion, }
 Phys.\ Lett.\ B {\bf 148}, 451 (1984).
 
\bibitem{Buscher:1987uw}
 T.~Buscher, U.~Lindstr\"om and M.~Ro\v{c}ek,
 {\em New supersymmetric sigma models with Wess-Zumino Terms},
 Phys.\ Lett.\ B {\bf 202},94 (1988).
 
\bibitem{Sevrin:1996jr}
 A.~Sevrin and J.~Troost,
 {\it Off-shell formulation of N = 2 nonlinear sigma-models},
 Nucl.\ Phys.\ B {\bf 492}, 623-646 (1997).
 [arXiv:hep-th/9610102].

\bibitem{Grisaru:1997pg}
 M.~T.~Grisaru, M.~Massar, A.~Sevrin and J.~Troost,
 {\it The quantum geometry of N = (2,2) nonlinear sigma-models},
 Phys.\ Lett.\ B {\bf 412}, 53 (1997)
 [arXiv:hep-th/9706218].
 
\bibitem{Grisaru:1997ep}
 M.~T.~Grisaru, M.~Massar, A.~Sevrin and J.~Troost,
 {\it Some aspects of N = (2,2), D = 2 supersymmetry},
 Fortsch.\ Phys.\ {\bf 47}, 301 (1999)
 [arXiv:hep-th/9801080].
 
\bibitem{Bogaerts:1999jc}
 J.~Bogaerts, A.~Sevrin, S.~van der Loo and S.~Van Gils,
 {\it Properties of semi-chiral superfields},
 Nucl.\ Phys.\ B {\bf 562}, 277 (1999)
 [arXiv:hep-th/9905141].
 
\bibitem{Gualtieri:2003dx}
M. Gualtieri, 
{\it Generalized complex geometry}, Oxford University 
DPhil thesis [arXiv:math.DG/0401221]; 
{\em Generalized complex geometry}, [arXiv:math.DG/0703298].

 \bibitem{hitchinCY}
N.~Hitchin,
 {\em Generalized Calabi-Yau manifolds,} Q. J. Math. 
 {\bf 54}, no. 3, 281-308 (2003) [arXiv:math.DG/0209099].
 
\bibitem{Lindstrom:2004eh}
 U.~Lindstr\"om,
 {\em Generalized N = (2,2) supersymmetric nonlinear sigma models,}
 Phys.\ Lett.\ B {\bf 587}, 216 (2004)
 [arXiv:hep-th/0401100].
 
\bibitem{Lindstrom:2004iw}
 U.~Lindstr\"om, R.~Minasian, A.~Tomasiello and M.~Zabzine,
 {\em Generalized complex manifolds and supersymmetry,}
 Commun.\ Math.\ Phys.\ {\bf 257}, 235 (2005)
 [arXiv:hep-th/0405085].
 
\bibitem{Bredthauer:2006hf}
 A.~Bredthauer, U.~Lindstr\"om, J.~Persson and M.~Zabzine,
 {\em Generalized Kaehler geometry from supersymmetric sigma models,}
 Lett.\ Math.\ Phys.\ {\bf 77}, 291 (2006)
 [arXiv:hep-th/0603130].
 
\bibitem{Lindstrom:2004hi}
 U.~Lindstr\"om, M.~Ro\v cek, R.~von Unge and M.~Zabzine,
 {\em Generalized Kaehler geometry and manifest N = (2,2) supersymmetric
 nonlinear sigma-models,}
 JHEP {\bf 0507}, 067 (2005)
 [arXiv:hep-th/0411186].
 
\bibitem{Lindstrom:2007qf}
 U.~Lindstr\"om, M.~Ro\v cek, R.~von Unge and M.~Zabzine,
 {\em Linearizing Generalized Kahler Geometry,}
 JHEP {\bf 0704}, 061 (2007)
 [arXiv:hep-th/0702126].
 
\bibitem{Lindstrom:2005zr}
 U.~Lindstr\"om, M.~Ro\v{c}ek, R.~von Unge and M.~Zabzine,
{\em Generalized K\"ahler manifolds and off-shell supersymmetry},
 Commun.\ Math.\ Phys.\ {\bf 269}, 833 (2007)
 [arXiv:hep-th/0512164].

\bibitem{Lindstrom:2007xv}
 U.~Lindstr\"om, M.~Ro\v cek, R.~von Unge and M.~Zabzine,
 {\em A potential for generalized Kaehler geometry,}
 To be published in {\bf Handbook of pseudo-Riemannian Geometry and Supersymmetry}
 [arXiv:hep-th/0703111].
 
\bibitem{Bredthauer:2006sz}
 A.~Bredthauer,
 {\em Generalized hyperkaehler geometry and supersymmetry,}
 Nucl.\ Phys.\ B {\bf 773}, 172 (2007)
 [arXiv:hep-th/0608114].
 
\bibitem{Bobby}
 B.~Ezhuthachan and D.~Ghoshal,
 {\it Generalised hyperK\"ahler manifolds in string theory},
 JHEP {\bf 04}, 083 (2007).

\bibitem{Goteman:2009xb}
 M.~G\"oteman and U.~Lindstr\"om,
 {\em Pseudo-hyperkahler Geometry and Generalized Kahler Geometry,}
 arXiv:0903.2376 [hep-th].

\bibitem{Hull:1985pq}
 C.~M.~Hull, A.~Karlhede, U.~Lindstr\"om and M.~Ro\v cek,
 {\it Nonlinear sigma models and their gauging in and out of superspace},
 Nucl.\ Phys.\ B {\bf 266}, 1 (1986).

\bibitem{Lindstrom:1994mw}
 U.~Lindstr\"om, I.~T.~Ivanov and M.~Ro\v cek,
 {\em New N=4 superfields and sigma models},
 Phys.\ Lett.\ B {\bf 328}, 49 (1994)
 [arXiv:hep-th/9401091].

\bibitem{Hull:2008vw}
 C.~M.~Hull, U.~Lindstr\"om, M.~Ro\v cek, R.~von Unge and M.~Zabzine,
 {\em Generalized Kahler geometry and gerbes},
 JHEP {\bf 0910}, 062 (2009),
 arXiv:0811.3615 [hep-th].
 
 \bibitem{YanoAko}
 K.~Yano and M.~Ako,
 {\it On certain operators associate with tensor fields},
 Kodai. Math. Sem. Rep. {\bf 20}, 414 (1968).
 
 \bibitem{MagriMorosi}
 F.~Magri and C.~Morosi, 
 {\it A geometric characterization of integrable 
 Hamiltonian systems through the theory of Poisson-Nijenhuis manifolds},
 Universit\`a di Milano, Quaderno S 19 (1984).
 
\bibitem{Howe:1988cj}
 P.~S.~Howe and G.~Papadopoulos,
 {\em Further remarks on the geometry of two-dimensional 
 nonlinear sigma models}, 
 Class.\ Quant.\ Grav.\ {\bf 5}, 1647 (1988).
 
\bibitem{Yano:1961}
 K.~Yano,
 {\it On a structure f satisfying $f^3+f=0$}, Tech. Rep. Univ. of Washington, {\bf 12} (1961); 
 {\it On a structure defined by a tensor field of type $(1,1)$ satisfying $f^3+f=0$},
 Tensor, N.S. {\bf 14}, 9 (1963).
 
\bibitem{IshiharaYano}
 S. Ishihara and K. Yano,
 {\it On integrability conditions of a structure $f$ satisfying $f^3+f=0$},
 Quart. J. Math. {\bf 15}, 217-222 (1964).


\end{thebibliography}
\end{document}